\documentclass{raa}            % referee version: for submission

\usepackage{graphicx,times}             %for PS/EPS graphics inclusion, new
\usepackage{natbib}
\usepackage{amssymb,amsmath}
\usepackage{xcolor}
\bibpunct{(}{)}{;}{a}{}{,}

\usepackage[a4paper=true,dvipdfm=true,pagebackref=true]{hyperref}
\hypersetup{colorlinks = true, linkcolor = green, anchorcolor = red, citecolor = blue, filecolor = red, pagecolor = red, urlcolor = red}

\begin{document}

   \title{Can we constrain the evolution of HI bias using configuration entropy?
%\footnotetext{$*$ Supported by the National Natural Science Foundation of China.}
}
%   \subtitle{I. Place Your Subtitle Here}

   \volnopage{Vol.0 (20xx) No.0, 000--000}      %%preserved for Editor. DOn't remove!
   \setcounter{page}{1}          %%starting page, preserved for Editor. DOn't remove!

   \author{Biswajit Das
      \inst{1}
   \and Biswajit Pandey
      \inst{2}
   %\and B. J. Smith
     % \inst{3}
   }

   \institute{Department of Physics, Visva-Bharati University, Santiniketan, 
              Birbhum, 731235, India
             {\it bishoophy@gmail.com}\\
        \and
	     {Departmnet of Physics, Visva-Bharati University, Santiniketan, 
	     Birbhum, 731235, India {\it biswap@visva-bharati.ac.in}}\\
        %\and
             %Full institute address for the third author\\
\vs\no
   {\small Received~~20xx month day; accepted~~20xx~~month day}}

\abstract{We study the evolution of the configuration entropy of HI
  distribution in the post-reionization era assuming
  different time evolution of HI bias. We describe time evolution of
  linear bias of HI distribution using a simple form $b(a)=b_{0}
  a^{n}$ with different index $n$. The derivative of the configuration
  entropy rate is known to exhibit a peak at the scale factor
  corresponding to the $\Lambda$-matter equality in the unbiased
  $\Lambda$CDM model. We show that in the $\Lambda$CDM model with
  time-dependent linear bias, the peak shifts to smaller scale factors
  for negative values of $n$. This is related to the fact that the
  growth of structures in the HI density field can significantly slow
  down even before the onset of $\Lambda$ domination in presence of a
  strong time evolution of the HI bias. We find that the shift is
  linearly related to the index $n$. We obtain the best fit relation
  between these two parameters and propose that identifying the
  location of this peak from observations would allow us to constrain
  the time evolution of HI bias within the framework of the
  $\Lambda$CDM model.   }

   \authorrunning{B. Das \& B. Pandey}            %author_head in even pages
   \titlerunning{HI bias from entropy}  % title_head in odd pages

   \maketitle
%% The author head (on even pages) and the title head (on odd pages) will be
%% automatically extracted from \author{} and \title{}. Whenever the title is too long,
%% you will be asked to supply a shorter one by inserting either \authorrunning{} or
%% \titlerunning{} before \maketitle. Anyway, you can specify your own heads.
%%
%%
%% Note: In the following text body of your manuscript, please note several differences from
%%       other major journals:
%% (1) \subsection{Please Capitalize the First Letter of Each Notional Word in Subsection Title}
%% (2) Please Capitalize the First Letter of Each Notional Word in all tables' captions

%
%________________________________________________ sections below
%
\section{Introduction}           %% first-level sections will be auto-capitalized
\label{sect:intro}

Our knowledge about the present day galaxy distribution in the nearby
Universe has been revolutionized by the modern galaxy surveys (SDSS,
\citealt{york}; 2dFGRS, \citealt{colles}; 2MRS, \citealt{huchra})
carried out over the last few decades. Many cosmological observations
suggest that most of the mass in the Universe is in the form of an
unseen dark matter which is yet to be directly detected by
observations. The galaxies are known to be a biased tracer of the
underlying dark matter distribution. On large scales, it is believed
that the fluctuations in the galaxy distribution and the dark matter
distribution are linearly related by a bias parameter \citep{kaiser84,
  dekel}. The linear bias parameter is known to be scale-independent
on large scales \citep{mann} but is expected to evolve with time
\citep{fry,tegmark}.  The time evolution of the linear bias parameter
determine the evolution of the large scale distribution of the tracer
relative to the underlying mass distribution. However the galaxies
have not always been in place. They are the product of the non-linear
evolution of the cosmic density field. Thanks to the improvement of
computing power and algorithms, modern day N-body simulations
\citep{springel, vogelsberger} can give us a clear idea about the
emergence of structures through non-linear evolution. In fact, the
understanding of the process of structure formation has become so good
that it has become a standard tool for testing cosmological
models. 

Early measurements of the two point correlation function for galaxies
and galaxy clusters did not match, indicating that both cannot be
unbiased tracers of the underlying matter distribution
\citep{kaiser84}. Various statistical tools are used to measure the
linear bias parameter from observations. One can employ the two-point
correlation function and power spectrum to determine the linear bias
parameter \citep{nor,teg,zehavi10}. The redshift space distortions of
the two-point correlation function and power spectrum \citep{kaiser87,
  hamilton} can be also employed to measure the linear bias parameter
\citep{haw, teg}. The other alternatives which have been successfully
used to compute the linear bias parameter are the three-point
correlation function and bispectrum \citep{feldman,verde,gaztanaga},
filamentarity \citep{pandey07} and information entropy
\citep{pandey17a}. It has been shown by
  \citet{pandey17a} that measurement of bias using information entropy
  requires only $O(N)$ operations as compared to $O(N^2)$ or at least
  $O(N \log N)$ operations required by the two-point correlation
  function and the power spectrum.

Galaxies do not exist at high redshift whereas the neutral Hydrogen
(HI) is present throughout the history of the Universe since its
formation after the recombination at $z \sim 1100$. The redshifted 21
cm line from neutral Hydrogen would reveal a wealth of information
about the formation and evolution of structures in the Universe. A
number of surveys (HIPASS, \citealt{zwaan}; ALFALFA, \citealt{martin})
have been designed to map the HI content of galaxies in the nearby
Universe. A significant effort has been also directed to detect the
redshifted 21 cm signal using different ongoing and upcoming radio
interferometric facilities (GMRT, \citealt{paciga}; LOFAR,
\citealt{vanharlem}; MWA \citealt{bowman}; SKA,
\citealt{mallema}). The redshifted 21 cm line can be used as a
promising probe of the large scale structures over a wide redshift
range \citep{bharad01, bharadsethi}. The knowledge about the HI bias
and its time evolution is also important in understanding the
uncertainties associated with the measured intensity fluctuation power
spectrum. Several studies have been carried out to measure the HI bias
\citep{martin, masui, switzer} at low redshifts ($z<1$) but presently
the evolution of HI bias with redshift is not known. Some theoretical
and observational constraints on the evolution of HI bias over the
redshift range $0-3.5$ has been discussed in \citealt{hamsa} and
references therein.

Most of the HI resides in the intergalactic medium
  during the epoch of reionization. The HI distribution deviates from
  the dark matter distribution due to the non-linear growth of ionized
  hydrogen (HII) bubbles and formation of early galaxies during this
  epoch. The HI distribution can not be treated as a tracer of the
  underlying matter density field during the epoch of
  reionization. But most of the HI settles in halos after reionization
  and the HI distribution can be treated as a reliable tracer of the
  total mass distribution in the post-reionization era. The HI bias in
  the post-reionization era has been studied in some works
  \citep{bagla,sarkar} by populating HI in dark matter halos from
  N-body simulations.

Recently, it has been suggested that the measurement of the
configuration entropy \citep{pandey1, pandey3} of the mass
distribution in the Universe can be used to test the different
cosmological models \citep{das1}, determine the mass density parameter
and cosmological constant \citep{pandey2} and constrain the dark
energy equation of state parameters \citep{das2}. In
  the present work, we propose a theoretical framework based on the
  study of configuration entropy which may allow us to probe the
  evolution of HI bias in the post-reionization era from future
  redshifted 21 cm observations.

%% Authors can give a citation as 'Michel et al. 1992'.
%% You may also use \cite, \citep and \citet for citation, and use Table~1 or Figure~1
%% and so forth. Using \ref and \label for cross-references of Tables/Figures
%% is a good way in adjusting/adding/removing text, tables or figures.

\section{Theory}
\subsection{Evolution of configuration entropy}
 We consider the HI distribution in the
  post-reionization era which can be treated as a biased tracer of the
  underlying dark matter distribution. We are interested in studying
  the time evolution of the linear bias of HI distribution using
  configuration entropy. Let us consider a large comoving volume $V$
of the Universe and divide it into sub-volumes $dV$. Let the density
of HI in each of these sub-volumes at time $t$ be $\rho_{HI} (\vec{x},
t)$ where $\vec {x}$ is the comoving coordinate of the sub-volume
defined with respect to an arbitrary origin. The configuration entropy
of the HI density field can be defined as \citep{pandey1},
\begin{eqnarray}
S_c(t) = - \int \rho_{HI}(\vec{x}, t) \log \rho_{HI} (\vec{x}, t)\, dV.
\label{eq:one}
\end{eqnarray}
The definition of configuration entropy is motivated from the
definition of information entropy \citep{shannon48}.

The mass distribution of the Universe is often treated as an ideal
fluid to a good approximation. The continuity equation of this fluid
in an expanding Universe can be written as,
\begin{eqnarray}
\frac{\partial \rho_{HI}}{\partial t} + 3 \frac{\dot a}{a} \rho_{HI} +
\frac{1}{a} \nabla \cdot (\rho_{HI} \vec {v_{HI}}) = 0.
\label{eq:two}
\end{eqnarray}
In \autoref{eq:two}, $a$ is the cosmological scale factor and
$\vec{v_{HI}}$ is the peculiar velocity of the HI mass element. We can
combine \autoref{eq:one} and \autoref{eq:two} to get,
\begin{eqnarray}
\frac{dS_c(t)}{dt} + 3 \frac{\dot a}{a} S_c(t) - \frac{1}{a} \int
\rho_{HI} (3 \dot a + \nabla \cdot \vec {v_{HI}})\, dV = 0.
\label{eq:three}
\end{eqnarray}
We rewrite \autoref{eq:three} as, 
\begin{eqnarray}
\frac{dS_c(a)}{da} \dot a + 3 \frac{\dot a}{a} S_c(a) - 3 \frac{\dot
  a}{a} \int \rho_{HI}(\vec {x}, a)\, dV \nonumber \\ - \frac{1}{a} \int
\rho_{HI}(\vec {x}, a) \nabla \cdot \vec {v_{HI}}\, dV = 0,
\label{eq:four}
\end{eqnarray}
where the variable of differentiation has been changed from $t$ to
$a$. Here $\int \rho_{HI}(\vec {x}, a)\, dV = M_{HI}$ is the total
mass of HI contained inside the comoving volume $V$. The density of HI
at comoving location $\vec{x}$ can be expressed as $\rho_{HI}(\vec {x},
a) = \bar \rho_{HI} (1 + \delta_{HI} (\vec {x}, a))$, where $\delta_{HI}(\vec
{x}, a)$ is the density contrast at location $\vec{x}$ and $\bar
\rho_{HI} = \frac{M_{HI}}{V}$ is the average density of HI. In linear
perturbation theory, one can write $\delta_m(\vec {x}, a) = D(a)
\delta_m (\vec {x})$ and $\nabla \cdot \vec {v_{HI}} = - a \frac{\partial
  \delta_{HI}(\vec{x}, a)}{\partial t}$. Here, $D(a)$ is the growing mode
and $\delta_m(\vec{x})$ is the initial mass density perturbation at
location $\vec{x}$. We simplify \autoref{eq:four} using these
relations to get,
\begin{eqnarray}
\frac{dS_c(a)}{da} \dot a + 3 \frac{\dot a}{a} (S_c(a) - M_{HI}) 
- \frac{\bar \rho_{HI}}{a} \int \nabla \cdot \vec{v_{HI}}\, dV \nonumber \\
- \frac{\bar \rho_{HI}}{a} \int \delta_{HI}(\vec{x}, a) \nabla \cdot \vec{v_{HI}}\, dV = 0.
\label{eq:five}
\end{eqnarray}

In the linear bias assumption,
\begin{eqnarray}
\delta_{HI} (\vec{x}, a) = b(a) \delta_m (\vec{x}, a),
\label{eq:six}
\end{eqnarray}
where $b(a)$ is the scale-independent linear bias parameter and
$\delta_{HI} (\vec{x}, a)$ and $\delta_m (\vec{x}, a)$ are the density
contrast corresponding to HI and the underlying mass density
field respectively. So,
\begin{eqnarray}
\nabla \cdot \vec{v_{HI}} = - a \dot a \left[D(a)
  \frac{db(a)}{da} + b(a) \frac{dD(a)}{da}\right] \delta_m (\vec{x}).
\label{eq:seven}
\end{eqnarray}
We combine \autoref{eq:seven} and \autoref{eq:five} and simplify to
get,
\begin{eqnarray}
\frac{dS_c(a)}{da} + \frac{3}{a} (S_c(a) - M_{HI}) + \bar \rho_{HI} B(a) \int
\delta^2_m(\vec{x})\, dV = 0.
\label{eq:eight}
\end{eqnarray}
Here, $B(a) = b(a)D(a)\left[D(a)\frac{db(a)}{da} +
  b(a)\frac{f(a)D(a)}{a}\right]$ where $f(a) = \frac{a}{D(a)}
\frac{dD(a)}{da}$ is the dimensionless linear growth rate.

This equation governs the evolution of configuration entropy of the HI
distribution in presence of time evolution of HI bias. One can
integrate \autoref{eq:eight} to get
\begin{eqnarray}
\frac{S_c(a)}{S_c(a_i)} = \frac{M_{HI}}{S_c(a_i)} + \left[1 -
  \frac{M_{HI}}{S_c(a_i)}\right]\Big(\frac{a_i}{a}\Big)^3 \nonumber \\- \Bigg(\frac{\bar \rho_{HI} \int
  \delta^2_m(\vec{x})\,dV}{S_c(a_i)a^3}\Bigg) \int_{a_i}^a da^{\prime}
a^{\prime 3} B(a^{\prime}).
\label{eq:nine}
\end{eqnarray}
Here $a_i$ is some initial scale factor and $S_c(a_i)$ is the initial
configuration entropy. In our analysis we have chosen $a_i =0.05$.

We find the evolution of the ratio of configuration entropy to its
initial value by numerically calculating the integral in the third
term for different time evolution of bias and substituting back at
\autoref{eq:nine}. We set the product $\bar \rho_{HI} \int
\delta^2_m(\vec{x})\,dV = 1$ for simplicity. The choices of $S_c(a_i)$
and $M_{HI}$ are arbitrary and in no way depend on the cosmological
model concerned. Choosing $S_c(a_i)>M_{HI}$ or $S_c(a_i)<M_{HI}$
causes a sudden growth or decay in $\frac{S_c(a)}{S_c(a_i)}$ near the
initial scale factor $a_i$, respectively. We have chosen $S_c(a_i) =
M_{HI}$ in our analysis to ignore these transients caused by the
initial conditions. The integral in the third term of
\autoref{eq:nine} involves evolution of growing mode, time dependent
bias and their derivatives. We describe these in detail in the next
two subsections.

\begin{figure*}
\resizebox{!}{6cm}{\includegraphics{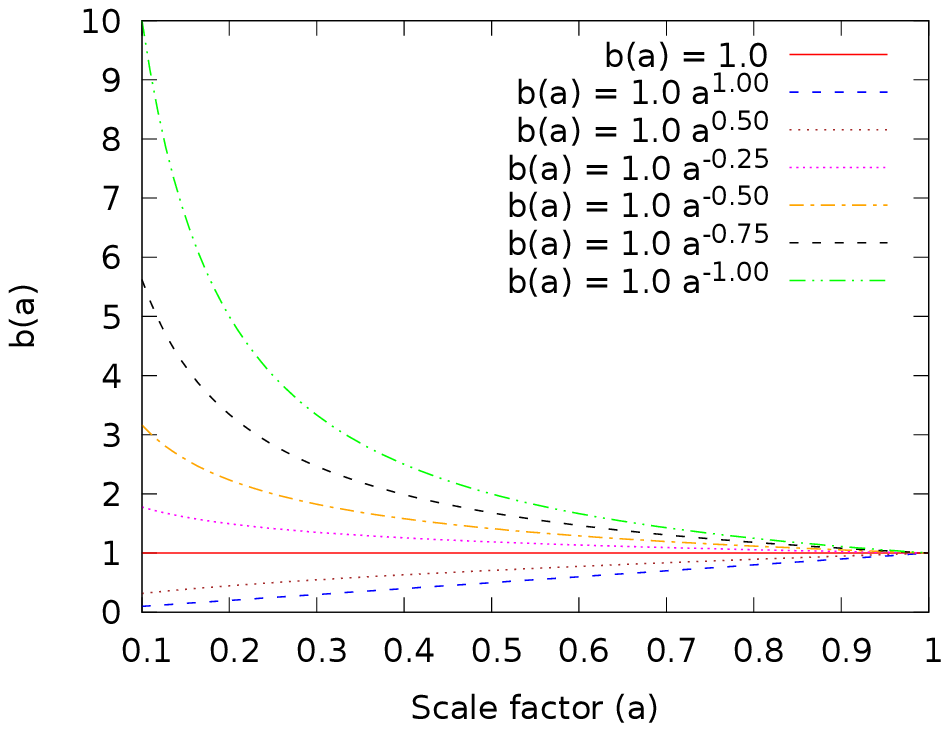}}
\resizebox{!}{6cm}{\includegraphics{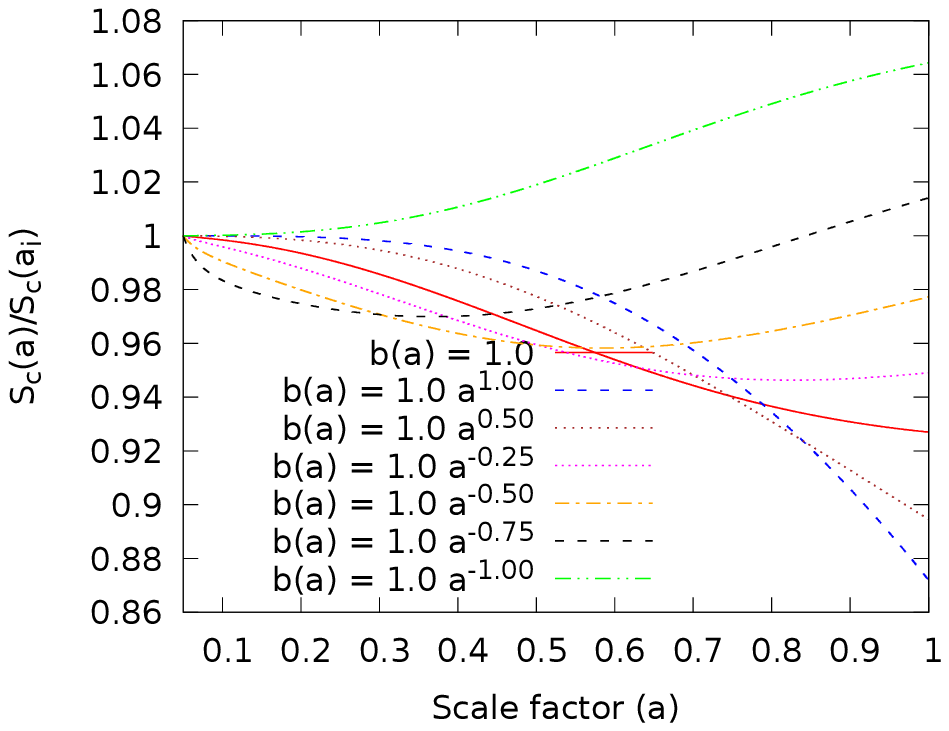}}\\
\resizebox{!}{6cm}{\includegraphics{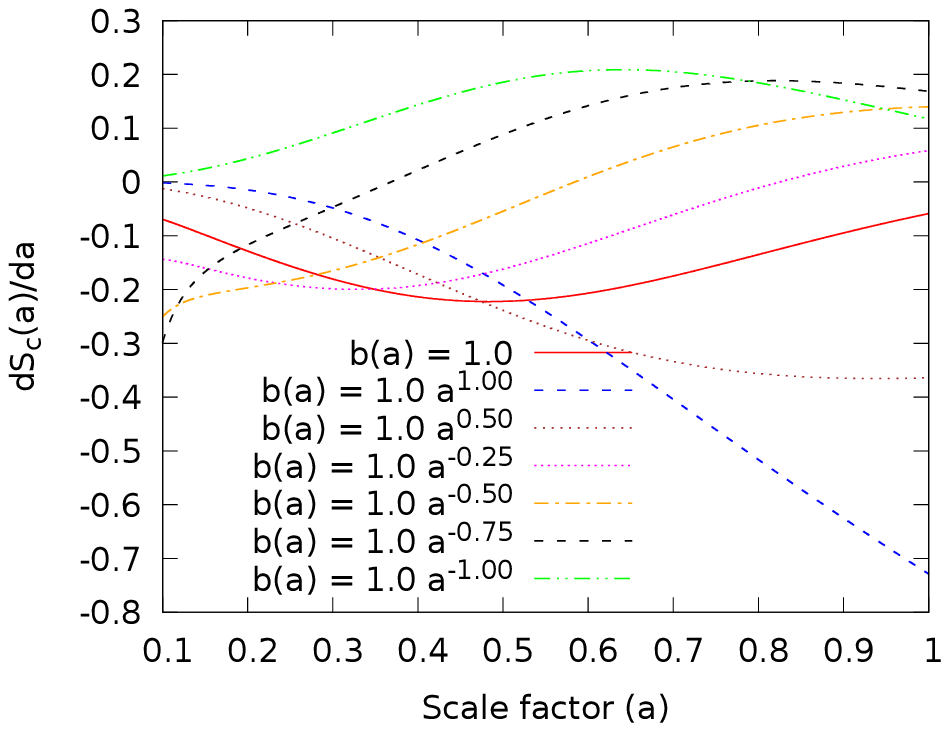}}
\resizebox{!}{6cm}{\includegraphics{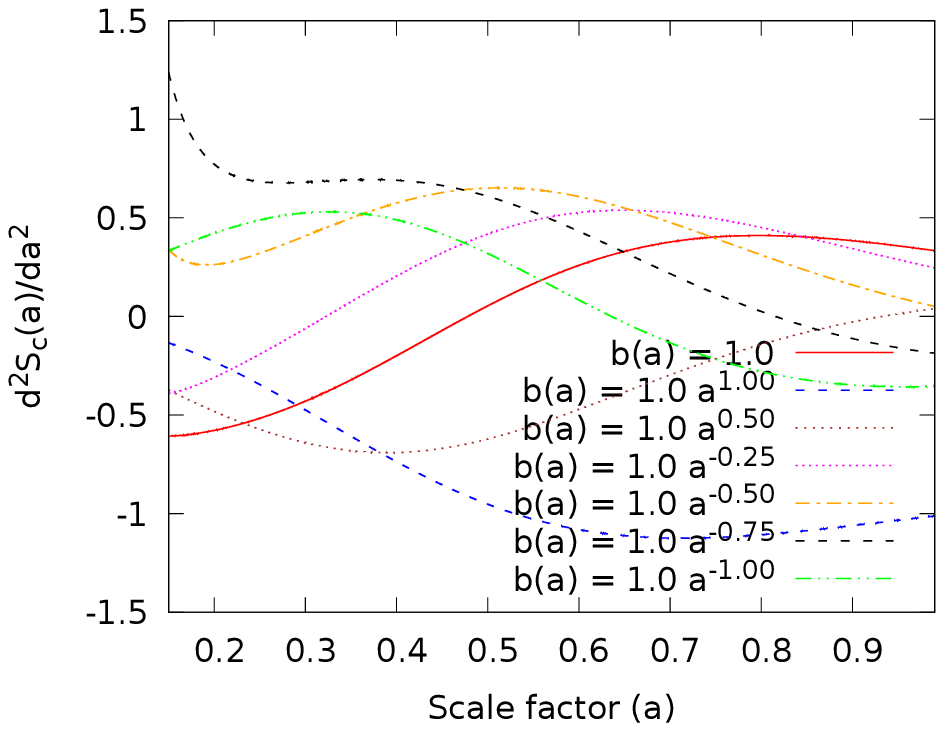}}
\caption{The top left panel of the figure shows the evolution of bias
  with scale factor for different models. The top right panel shows
  the evolution of $S_c(a)/S_c(a_i)$ with scale factor for different
  evolution of bias within the $\Lambda$CDM model. The bottom left and
  right panel respectively show the evolution of $dS_c(a)/da$ and
  $d^2S_c(a)/da^2$ with scale factor for the same models. The results
  for the unbiased case $b=1$ are also shown in each panel for
  comparison.}
\label{fig:one}
\end{figure*}

%
%               one-column-spanning table
%________________________________________ Table 2: Use_of_the routines

%%Please Capitalize the First Letter of Each Notional Word in table's caption

%%%%%%%%%%%%%%%%%%%%%%%%%%%%%%%%%%%%%%%%%%%%%%%%%%%%%%%%%%%%%%
%%     Examples for figures using graphicx for LaTeX 2e
%%               -- our recommended way for embodying graphics
%%%%%%%%%%%%%%%%%%%%%%%%%%%%%%%%%%%%%%%%%%%%%%%%%%%%%%%%%%%%%%
%
%      A figure as large as the width of the column
%-------------------------------------------------------------
   
\subsection{Growth rate of density perturbations}

The CMBR observations suggest that the Universe was highly isotropic
at $z \sim 1100$. But the present day Universe is not homogeneous and
isotropic on small scales. We find galaxies and clusters of galaxies
where huge mass is accumulated over a small region whereas there are
large empty regions or voids with very little amount of mass. The
linear perturbation theory provides a theoretical framework to
understand the growth of structures from tiny fluctuations seeded in a
homogeneous and isotropic distribution in the early Universe. In the
currently accepted paradigm, gravitational instability is the primary
mechanism behind the formation of structures in the Universe. CMBR
observations indicate that inhomogenities of very small magnitude were
present in the matter distribution at the time of recombination. These
tiny inhomogeneities get amplified by the gravitational instability
over time. When the density contrast is much smaller than $1$, its
evolution can be described by the following differential equation,
\begin{eqnarray}
\frac{\partial^2 \delta_m(\vec{x}, t)}{\partial t^2} + 2 H(a) \frac{\partial \delta_m(\vec{x}, t)}{\partial t} - \frac{3}{2} \Omega_{m0} H^2_0 \frac{1}{a^3} \delta_m(\vec{x}, t) = 0.
\label{eq:ten}
\end{eqnarray}
Here we have considered perturbation to only matter
component. $\Omega_{m0}$ and $H_0$ are the present value of density
parameter for matter and Hubble parameter, respectively. This equation
governs the growth of density perturbation in the underlying matter
distribution. The equation has a growing mode solution of the form
$\delta_m(\vec{x}, t) = D(t)\delta_m(\vec{x})$. The growing mode
solution can be expressed as \citep{peebles}
\begin{eqnarray}
D(a) = \frac{5}{2} \Omega_{m0} X^{\frac{1}{2}}(a) \int_0^a \frac{da^{\prime}}{a^{\prime 3} X^{\frac{3}{2}}(a^{\prime})},
\label{eq:eleven}
\end{eqnarray}
where $X(a) = \frac{H^2(a)}{H_0^2} = [\Omega_{m0} a^{-3} +
  \Omega_{\Lambda0}]$. Here $\Omega_{\Lambda0}$ is the present value
of the density parameter corresponding to cosmological constant.

The dimensionless linear growth rate $f(a) = \frac{d\log D(a)}{d \log
  a}$ in a universe with no curvature can be approximated as
\citep{lahav}
\begin{eqnarray}
	f(a) = \Omega_m(a)^{0.6} + \frac{1}{70}\left[1 - \frac{1}{2}
          \Omega_m(a) (1 + \Omega_m(a))\right].
\label{eq:twelve}
\end{eqnarray}
Here $\Omega_m(a) = \frac{\Omega_{m0} a^{-3}}{X(a)}$. We have used
$\Omega_{m0} = 0.3$ and $\Omega_{\Lambda0} = 0.7$ throughout the
present work.

\begin{figure*}
\resizebox{!}{6cm}{\includegraphics{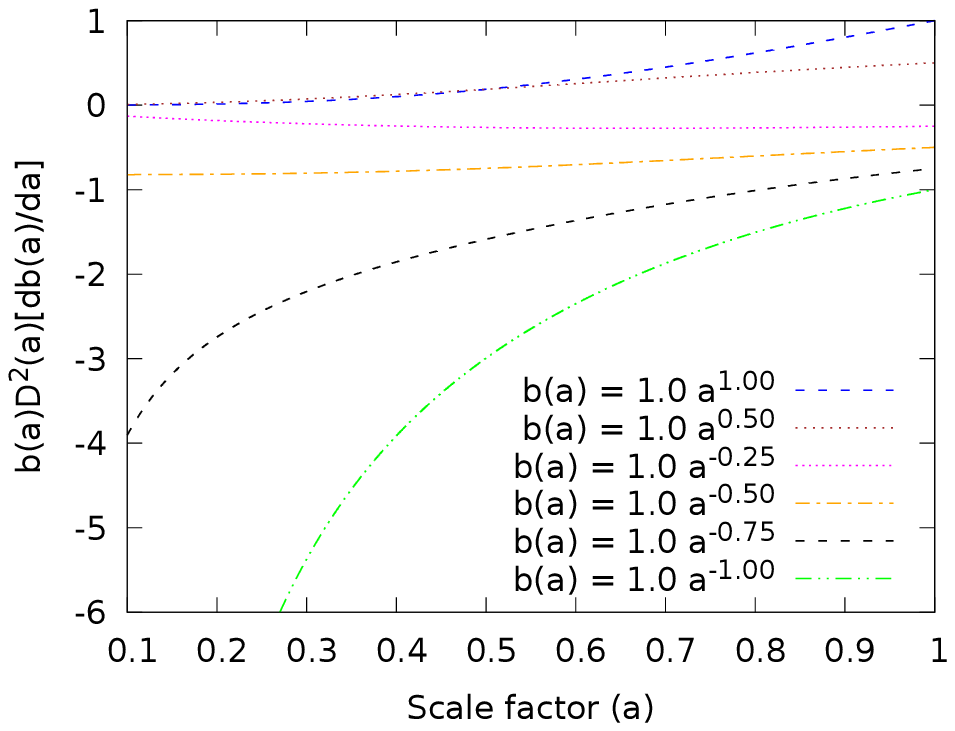}}
\resizebox{!}{6cm}{\includegraphics{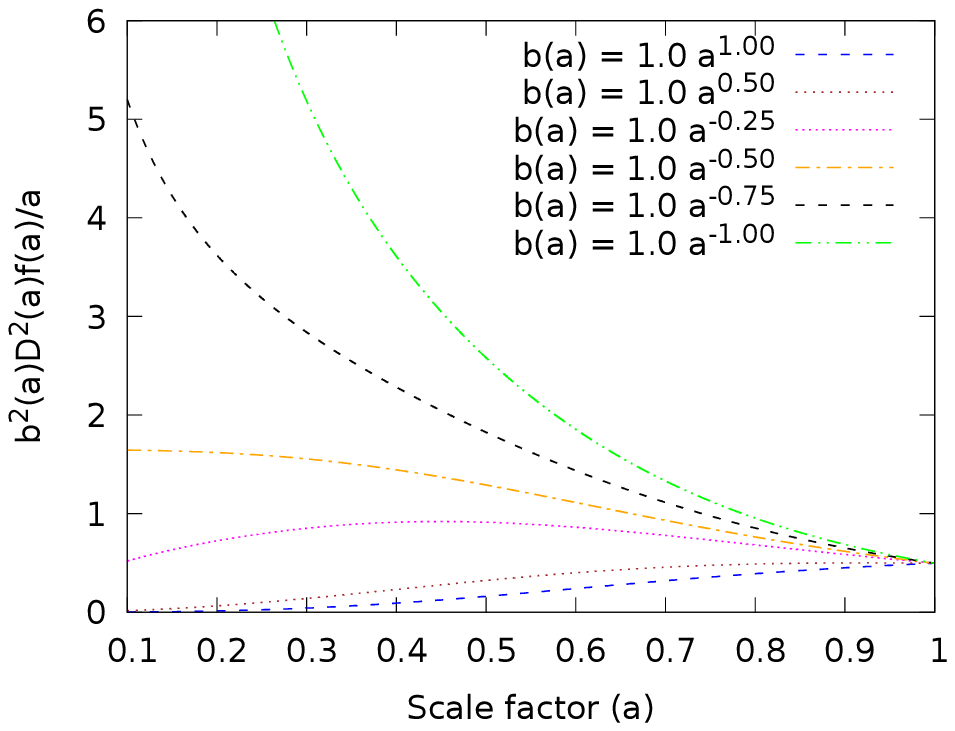}}
\caption{The left panel shows the evolution of the first term in
  $B(a)$ with scale factor for models with different time evolution of
  bias. The right panel shows the evolution of the second term in
  $B(a)$ for the same models.}
\label{fig:two}
\end{figure*}

\begin{figure*}
\resizebox{!}{6cm}{\includegraphics{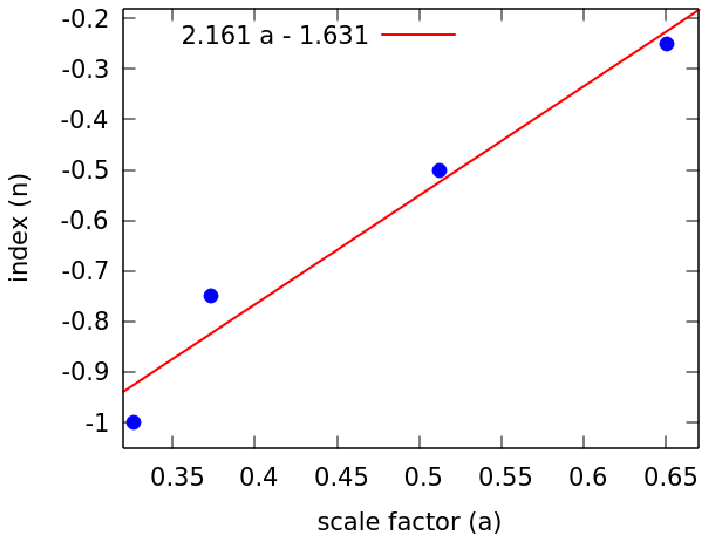}}
\resizebox{!}{6cm}{\includegraphics{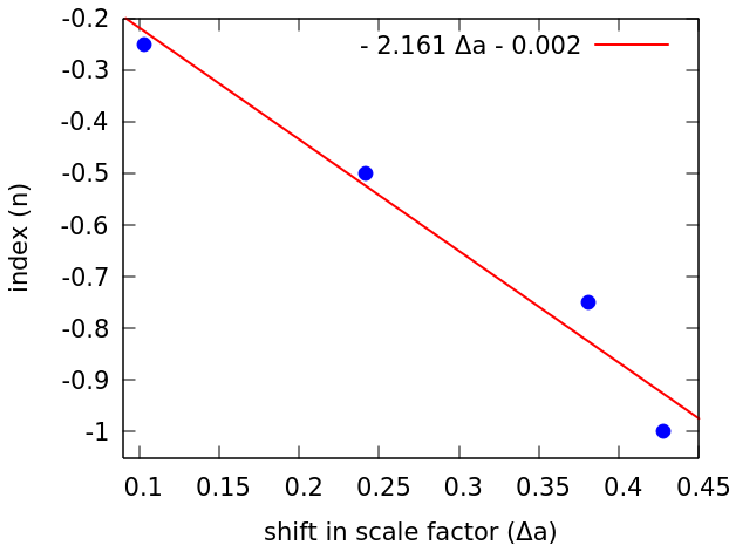}}
\caption{The left panel shows the index (n) as a function
  of the location of the peaks in the derivative of entropy rate. The right
  panel shows the index (n) as a function of the shift of peak in the derivative   of entropy rate with respect to unbiased
  $\Lambda$CDM model. We show together the
  best fit straight lines in both the panels.}
\label{fig:three}
\end{figure*}

\subsection{Evolution of HI bias}
The time evolution of the HI bias parameter is expected to affect the
time evolution of the configuration entropy of the HI density
field. We consider a simple power law of the form $b(a)=b_{0} a^{n}$
with different possible values of $n$. The functional form is
motivated by \citet{bagla} where $b(z) \propto (1 + z)^{0.5}$ was
reported to give a reasonably good description of the evolution of HI
bias in the simulated HI distributions from the N-body simulations.
We consider the following values of $n$ in our analysis: $n = -1,
-0.75, -0.5, -0.25, 0.5, 1$.  We also incorporate the unbiased
$\Lambda$CDM model in this framework by putting $b(a) = b_0$. We set
$b_0 = 1$ in all the models considered here.

\section{Results and Conclusions}

We show the evolution of the HI bias with scale factor for different
values of $n$ in the top left panel of \autoref{fig:one}. The
amplitude of the bias at any given scale factor depends on the index
$n$. The HI bias monotonically decreases with increasing scale factor
for negative $n$. A negative value of $n$ indicates that the HI
density field was strongly biased in the past which decreases with
time and eventually reaches unity at present. The decrease in bias
corresponds to an overall dilution in the clustering of the HI mass
distribution. The evolution of $\frac{S_c(a)}{S_c(a_i)}$ with scale
factor for all these models is shown in the top right panel of
\autoref{fig:one}. The evolution of the configuration entropy is
primarily governed by the growth of density perturbations which in
turn is affected by the dynamics of the expansion of the
Universe. Expansion of the Universe slows down the growth of
perturbations. Besides the expansion, the time evolution of bias would
also play an important role in controlling the dissipation of the
configuration entropy of the Universe. For example, all the models
with negative $n$ show a decrease in the configuration entropy at
earlier times. However the dissipation slows down with time and in
some cases it may even reverse its behaviour and starts to grow again
with time. The time of reversal from dissipation to growth depends on
the index $n$. More negative index leads to an early reversal in the
behaviour of the configuration entropy.

The lower left panel of \autoref{fig:one} shows the entropy rate as a
function of scale factor in models with different $n$. The entropy
rate is decided by the function $B(a) =
b(a)D(a)\left[D(a)\frac{db(a)}{da} + b(a)\frac{f(a)D(a)}{a}\right]$
which consists of two terms and the combined contribution from these
two terms decides the behaviour of the entropy rate at any given time
for any specific model. The two terms are separately plotted as
function of the scale factor for different models in the left and
right panels of \autoref{fig:two}. Clearly a growth in entropy is
expected when $B(a)$ is negative and a positive $B(a)$ is associated
with entropy dissipation. For example $B(a)$ is negative at all scale
factor for $n=-1$ and this implies that there will be no dissipation
of entropy in this model. On the other hand the model with $n=1$ and
$n=0.5$ have positive $B(a)$ at all scale factors and there is a
continuous dissipation of entropy in these models. All the other
models considered here show dissipation of entropy at some
scale factors and growth of entropy at some other scale factors. A
zero up crossing in the entropy rate corresponds to a local minimum in
the configuration entropy. Clearly this zero up crossing appears at a
smaller scale factor for more negative values of $n$.

We show the derivative of the entropy rate in these models in the
lower right panel of \autoref{fig:one}. The derivative of the entropy
rate exhibits a peak in all the models with negative $n$.  We find
that the location of the peak is sensitive to the index $n$ and it
appears at a smaller scale factor for models with smaller index. In an
earlier work, \citet{pandey2} noted that in the unbiased $\Lambda$CDM
model, this peak is located at the scale factor corresponding to the
$\Lambda$-matter equality. We have used $\Omega_{m0}=0.3$ and
$\Omega_{\Lambda}=0.7$ in the $\Lambda$CDM model. So in the unbiased
$\Lambda$CDM model the peak is expected to appear at $a=0.754$. This
can be clearly seen in the result shown for the unbiased $\Lambda$CDM
model in the same panel. Now the location of this peak is shifted
towards a smaller scale factor when time evolution of bias is
considered within the $\Lambda$CDM model. The shift is measured with
reference to the location of the peak in the unbiased $\Lambda$CDM
model. The magnitude of the entropy rate slows down after the
occurrence of this peak. In the unbiased $\Lambda$CDM model, the
structure formation starts to slow down after the onset of $\Lambda$
domination. The bias models with negative value of $n$ dilute the
clustering and slows down the structure formation even before the
$\Lambda$-matter equality. This effect would manifest in a more
prominent way in the models with more negative $n$. So the peak in the
derivative of the entropy rate is expected to exhibit a larger shift
in these models.  We measure the location of the peak in the models
with different negative index and find them to be linearly
related. The location and the shift of the peak are shown as a
function of the index in the left and right panels of
\autoref{fig:three} respectively. The best fit relations between these
parameters are also shown in the same figure.

We also consider two positive values of $n$ in the time evolution of
HI bias. A positive value of $n$ indicates that the HI density field
is anti biased with respect to the underlying mass density field and
the bias slowly increases from a very small positive value to unity at
present. A decrease in anti biasing with time would enhance the
clustering of the HI leading to a continuous dissipation of the
configuration entropy. In these model entropy initially show a slower
decrease than that of $\Lambda$CDM model but then decrease quite
quickly in the later part. We do not observe the peak in the
derivative of the entropy rate in these models and they can be easily
distinguished from the models with negative values of $n$. These
models are not realistic and we consider them only for the sake of
completeness.

In this work, we calculate the evolution of the configuration entropy of 
HI distribution in the post-reionization era assuming different time evolution 
of HI bias. We consider the flat $\Lambda$CDM model as the benchmark model of the Universe and
within it consider the time evolution of HI bias as, $b(a)=b_{0}
a^{n}$ with different values of the index $n$. We show that the time
evolution of bias alters the position of the peak in the derivative of
the entropy rate. The peak shifts towards a smaller scale factor for
negative index and is absent when the index is positive. We find that
the shift is linearly related with the index $n$ and a larger shift is
observed for a smaller index. We find the best fit relation between
these two parameters and propose that identifying the location of this
peak from observations would allow us to constrain the time evolution
of bias within the framework of the $\Lambda$CDM model. We note here 
that the best fit line does not exactly pass through the points in each of the 
plots of \autoref{fig:three} even though the points we get are from theoretical calculations 
and hence exact. The reason for that is that the linear fit is used 
as a first approximation but it gives a pretty good fit. We also note that if any of the future
surveys provides us with a suitable data set such that our method can be applied 
for analysis, there may be error bars which may be as big as the difference between the 
fit and the actual points. So, the linear approximation can work well in that situation.

One may consider some other quantity of the form $\int f(\rho_{HI}) dV$ 
and get another equation which might be used to constrain the HI bias function. 
The natural question that one can then ask is : why use configuration entropy? 
Part of the answer may be found in the introduction where it has been mentioned that 
this quantity has previously been used to study different cosmologcal problems. 
The introduction also mentions that measurement of bias using configuration entropy is 
computationally advantegeous compared to other methods. It has previously been shown 
that in a flat $\Lambda$CDM universe with only matter and cosmological constant with 
scalar perturbation, the evolution of derivative of entropy rate with scale factor shows 
a distinct peak at a scale factor which is equal to the scale factor where matter-$\Lambda$ 
equality occurs in that particular model. We calculate the shifts of the scale factor 
of the peak for biased tracer from the scale factor of the peak for unbiased case 
and find its correlation with the indices of bias function. Since we are comparing 
the unbiased case with the biased case, we are compelled to use configuration entropy 
as the preferred quantity of analysis.

One can also measure the HI bias by comparing the
  two-point correlation function or power spectrum of the HI
  distribution with that for the underlying mass
  distribution. Combining these measurements at multiple redshifts
  would provide the time evolution of HI bias. However such an analysis
  would require the knowledge of the distributions of dark matter
  density field at different redshifts which can be obtained by using
  N-bdoy simulations. Contrary to this, the proposed method in this
  work does not require the knowledge of the underlying mass density
  field at any redshift. The evolution of HI bias can be solely
  determined from the nature of evolution of the configuration entropy
  of the HI distribution. This is a remarkable advantage offered by
  the proposed method. It may be noted here that we do use the evolution 
  of growing mode of dark matter to calculate entropy, but the evolution equation of 
  growing mode is obtained under very general assumptions such as existence of scalar perturbation in an expanding
  universe with presence of dark matter and cosmological constant with no interaction
  between dark matter and dark energy. Finally we conclude that the method presented
in this work provides an alternative method to constrain the evolution
of HI bias using configuration entropy.

\begin{acknowledgements}
The authors would like to thank an anonymous reviewer for useful comments 
and suggestions which have helped us improve the quality of the paper.
BP acknowledges financial support from the Science and Engineering
Research Board (SERB), Department of Science \& Technology (DST),
Government of India through the project EMR/2015/001037. BP would also
like to acknowledge IUCAA, Pune for providing support through the
associateship programme.
\end{acknowledgements}

\label{lastpage}

\end{document}